\documentclass[12pt,preprint]{aastex}

\shorttitle{Galactic metric and gravitational lensing in brane
world models} \shortauthors{Harko\& Cheng}
\begin{document}
\title{Galactic metric, dark radiation, dark pressure and gravitational lensing in brane world models}
\author{T.~Harko$^1$ and K.~S.~Cheng$^2$}
\affil{Department of Physics, The University of Hong Kong,
Pokfulam Road, Hong Kong, Hong Kong SAR, P. R. China}
\email{$^1$harko@hkucc.hku.hk, $^2$hrspksc@hkucc.hku.hk}

\begin{abstract}
In the braneworld scenario, the four dimensional effective
Einstein equation has extra terms which arise from the embedding
of the 3-brane in the bulk. These non-local effects, generated by
the free gravitational field of the bulk, may provide an
explanation for the dynamics of the neutral hydrogen clouds at
large distances from the galactic center, which is usually
explained by postulating the existence of the dark matter.
Starting from the observational fact of the constancy of the
galactic rotation curves in spiral galaxies, we obtain the exact
galactic metric in the flat rotation curves region in the brane
world scenario. The basic physical parameters describing the
non-local effects (dark radiation and dark pressure) are obtained,
and their behavior is considered in detail. Due to the presence of
the bulk effects, the flat rotation curves could extend several
hundred kpc. The limiting radius for which bulk effects are
important is estimated and compared with the numerical values of
the truncation parameter of the dark matter halos, obtained from
weak lensing observations. There is a relatively good agreement
between the predictions of the model and observations. The
deflection of photons passing through the region where the
rotation curves are flat is also considered and the bending angle
of light is computed. The bending angle predicted by the brane
world models is much larger than that predicted by standard
general relativistic and dark matter models. The angular radii of
the Einstein rings are obtained in the small angles approximation.
 The predictions of the brane world model for the tangential
shear are compared with the observational data obtained in the
weak lensing of galaxies in the Red-Sequence Cluster Survey.
Therefore the study of the light deflection by galaxies and the
gravitational lensing could discriminate between the different
dynamical laws proposed to model the motion of particles at the
galactic level and the standard dark matter models.
\end{abstract}


\section{Introduction}

The rotation curves for galaxies or galaxy clusters should,
according to Newton's gravitation theory, show a Keplerian
decrease with distance $r$ of the orbital rotational speed
$v_{tg}$ at the rim of the luminous matter, $v_{tg}^2\propto
M(r)/r$, where $M(r)$ is the dynamical mass. However, one observes
instead rather flat rotation curves \citep{Ru80,Bin87,Pe96,Bo01}.
Observations show that the rotational velocities increase near the
center of the galaxy and then remain nearly constant at a value of
$v_{tg\infty }\sim 200-300$ km/s. This leads to a general mass
profile $M(r)\approx rv_{tg\infty }^2/G$ \citep{Bin87}.
Consequently, the mass within a distance $r$ from the center of
the galaxy increases linearly with $r$, even at large distances,
where very little luminous matter can be detected.

This behavior of the galactic rotation curves is explained by
postulating the existence of some dark (invisible) matter,
distributed in a spherical halo around the galaxies. The dark
matter is assumed to be a cold, pressureless medium. There are
many possible candidates for dark matter, the most popular ones
being the weekly interacting massive particles (WIMP) (for a
recent review of the particle physics aspects of dark matter see
\citet{OvWe04}). Their interaction cross section with normal
baryonic matter, while extremely small, are expected to be
non-zero and we may expect to detect them directly. It has also
been suggested that the dark matter in the Universe might be
composed of superheavy particles, with mass $\geq 10^{10}$ GeV.
But observational results show the dark matter can be composed of
superheavy particles only if these interact weakly with normal
matter or if their mass is above $10^{15}$ GeV \citep{AlBa03}.
Scalar fields or other long range coherent fields coupled to
gravity have also intensively been used to model galactic dark
matter
\citep{Nu00,MaGu01,MiSc02,Ca021,Li02,Mb04,Mat04,Fu04,Le04,MiPe04,He04}.

However, despite more than 20 years of intense experimental and
observational effort, up to now no {\it non-gravitational}
evidence for dark matter has ever been found: no direct evidence
of it and no annihilation radiation from it. Moreover, accelerator
and reactor experiments do not support the physics (beyond the
standard model) on which the dark matter hypothesis is based.

Therefore, it seems that the possibility that Einstein's (and the
Newtonian) gravity breaks down at the scale of galaxies cannot be
excluded {\it a priori}. Several theoretical models, based on a
modification of Newton's law or of general relativity, have been
proposed to explain the behavior of the galactic rotation curves.
A modified gravitational potential of the form $\phi =-GM\left[
1+\alpha \exp \left( -r/r_{0}\right) \right] /\left( 1+\alpha
\right) r$, with $\alpha =-0.9$ and $r_{0}\approx 30$ kpc can
explain flat rotational curves for most of the galaxies
\citep{Sa84,Sa86}.

In an other model, called MOND, and proposed by Milgrom
\citep{Mi83,BeMi84,Mi02,Mi03}, the Poisson equation for the
gravitational potential $\nabla ^{2}\phi =4\pi G\rho $ is replaced
by an equation of the form $\nabla \left[ \mu \left( x\right)
\left( \left| \nabla \phi \right| /a_{0}\right) \right] =4\pi
G\rho $, where $a_{0}$ is a fixed constant and $\mu \left(
x\right) $ a function satisfying the conditions $\mu \left(
x\right) =x$ for $x<<1$ and $\mu \left( x\right) =1$ for $x>>1$.
The force law, giving the acceleration
$a$ of a test particle  becomes $a=a_{N}$ for $a_{N}>>a_{0}$ and $a=\sqrt{%
a_{N}a_{0}}$ for $a_{N}<<a_{0}$, where $a_{N}$ is the usual
Newtonian acceleration. The rotation curves of the galaxies are
predicted to be flat, and they can be calculated once the
distribution of the baryonic matter is known.  A relativistic MOND
inspired theory was developed by Bekenstein \citep{Be04,Be05}. In
this theory gravitation is mediated by a metric, a scalar field
and a 4-vector field, all three dynamical.

Alternative theoretical models to explain the galactic rotation
curves have been proposed by \citet{Ma93}, \citet{Ma97},
\citet{Mo96}, \citet{Mo} and \citet{Ro04}.

The detection by \citet{Br96} of weak, tangential distortion of
the images of cosmologically distant, faint galaxies due to
gravitational lensing by foreground galaxies has opened a new
possibility of testing the alternative theories of gravitation and
the dark matter hypothesis. Background galaxies are observed to be
tangentially aligned around foreground galaxies because of the
latter population's gravitational lensing effect. The angular
dependence of the shear signal is consistent with the hypothesis
that galaxies are dominated by approximately isothermal haloes,
but can also be explained by assuming some phenomenological
modifications of the effective (Newtonian) gravitational force at
large distances \citep{MoTu01}.

Galaxy-galaxy lensing could provide very powerful constraints for
the alternative gravitational theories (for a general review of
week lensing and its applications see \citet{Ho02}). The
measurements rely on averaging over many background sources and
the signal is only appreciable at large angular separations from
the foreground deflectors. Thus, in the absence of dark matter,
the foreground galaxies can be regarded as simple lenses, and the
observational data can be used to constrain the deflection law
directly. A weak-lensing mass reconstruction of the interacting
cluster 1E 0657-558 has been presented in \citet{Cl04}.

Weak lensing has been mainly used to discuss and constrain MOND
\citep{Ga02,Ho02,Cl04}. \citet{Ga02} used the cluster of galaxies
MS2137-23, which presents the most constrained lensing
configuration of gravitational images ever detected in a distant
cluster of galaxies to constrain MOND type models. According to
this analysis,  the MOND model is not compatible with the
observations. The need for much more baryons in the MOND model
than for the dark matter paradigm implies significant dynamical
differences between these models, which can be explored at very
large radial distance. Observational evidence against this model
was also reported in \cite{Cl04}. For the interacting cluster 1E
0657-558 the observed offsets of the lensing mass peaks from the
X-ray gas directly demonstrate the presence, and dominance, of
dark matter in this cluster. However, this proof of dark matter
existence holds true even under the assumption of MOND. Based on
the observed gravitational shear-optical light ratios and the mass
peak-X-ray gas offsets, the "dark matter" component in a MOND
regime, that is, the dynamical MOND mass $M_m$, related to the
dynamical Newtonian mass $M_d$ by the relation
$M_{m}(r)=M_{d}/\sqrt{1+\left( a_{0}/a\right) ^{2}}$, would have a
total mass that is at least equal to the baryonic mass of the
galactic system. The observed shear and derived mass of the subclump
are significantly higher than could be produced by an isothermal sphere
with a $212$ km/s.

Weak gravitational lensing is a very promising method for the
study of the galactic dark matter halos. Results of the studies of
weak lensing by galaxies have been published recently by
\citet{Ho03,Ho04}. Using weak lensing the flattening of the
galactic dark matter halos was observed. These results suggest
that the dark matter halos are rounder than the light
distribution, with a halo ellipticity of the order of $e_{{\rm
halo}}\approx 0.33$. The average mass profile around galaxies was
also studied in the framework of two halo models, the truncated
isothermal sphere model and the Navarro-Frenk-White model. From
the point of view of the alternative theories of gravitation, the
main implications of these results is that spherical halos are
excluded with $99\%$ confidence, since most of the alternative
theories predict an almost isotropic weak lensing signal, which is
not observed.

In a series of recent papers \citep{Ha03,Ma04,Ha04},  several
classes of conformally symmetric solutions of the static
gravitational field equations in the brane world scenario
\citep{RS99a,RS99b}, in which our Universe is identified to a
domain wall in a $5$-dimensional anti-de Sitter space-time, have
been obtained (for a review of brane world models see
\citet{Mar04}). The static vacuum gravitational field equations on
the brane depend on the generally unknown Weyl stresses in the
bulk (a higher dimensional space-time, in which our universe is
embedded), which can be expressed in terms of two functions,
called the dark radiation $U$ and the dark pressure $P$ (the
projections of the Weyl curvature of the bulk, generating
non-local brane stresses) \citep{Ma04,Mar04,Ha04}. Generally, the
vacuum field equations on the brane can be reduced to a system of
two ordinary differential equations, which describe all the
geometric properties of the vacuum as functions of the dark
pressure and dark radiation terms \citep{Ha03}. In order to close
the system a functional relation between $U$ and $P$ is necessary.

In \citet{Ha03}, \citet{Ma04} and \citet{Ha04} the solutions of
the gravitational field equations on the brane have been obtained
by using some methods from Lie group theory. As a group of
admissible transformations the one-parameter group of conformal
motions has been considered. The main advantage of imposing
geometric self-similarity via a group of conformal motions is that
this condition also uniquely fixes the mathematical form of the
dark radiation and dark pressure terms, respectively, which
describe the non-local effects induced by the gravitational field
of the bulk. Thus there is no need to impose an arbitrary relation
between the dark radiation and the dark pressure.

As a possible physical application of the solutions of the
spherically symmetric gravitational field equations in the vacuum
on the brane the behavior of the angular velocity $v_{tg}$ of the
test particles in stable circular orbits has been considered
\citep{Ma04,Ha04}. The conformal factor $\psi $, together with two
constants of integration, uniquely determines the rotational
velocity of the particle. In the limit of large radial distances
and for a particular set of values of the integration constants
the angular velocity tends to a constant value. This behavior is
typical for massive particles (hydrogen clouds) outside galaxies
\citep{Bin87,Pe96,Bo01}, and is usually explained by postulating
the existence of the dark matter.

Thus, the rotational galactic curves can be naturally explained in
brane world models without introducing any additional hypothesis.
The galaxy is embedded in a modified, spherically symmetric
geometry, generated by the non-zero contribution of the Weyl
tensor from the bulk. The extra-terms, which can be described as
the dark radiation term $U$ and the dark pressure term $P$, act as
a "dark matter" distribution outside the galaxy. The existence of
the dark radiation term generates an equivalent mass term $M_U$,
which is linearly increasing with the distance, and proportional
to the baryonic mass of the galaxy, $M_{U}(r)\approx
M_{B}(r/r_{0})$. The particles moving in this geometry feel the
gravitational effects of $U$, which can be expressed in terms of
an equivalent mass. Moreover, the limiting value $v_{tg\infty }$
of the angular velocity can be obtained as function of the total
baryonic mass $M_B$ and radius $r_0$ of the galaxy as $v_{tg\infty
}\approx (2/\sqrt{3})\sqrt{GM_{B}/r_{0}}+ (1/12\sqrt{3})\left(
GM_{B}/r_{0}\right) ^{3/2}$ \citep{Ma04}.

For a galaxy with baryonic mass of the order $10^{9}M_{\odot }$
and radius of the order of $r_{0}\approx 70$ kpc, we have
$v_{tg\infty }\approx 287$ km/s, which is of the same order of
magnitude as the observed value of the angular velocity of the
galactic rotation curves. In the framework of this model all the
relevant physical parameters (metric tensor components, dark
radiation and dark pressure terms) can be obtained as functions of
the tangential velocity, and hence they can be determined
observationally.

However, imposing a group of conformal motions on the brane
implies a major restriction on the geometrical structure of the
space-time. Therefore it would be important to analyze the
behavior of the galactic rotation curves in the brane world model
without particular assumptions on the geometry of the brane. In
the framework of a Newtonian approximation of the brane metric the
possibility that the galactic rotation curves can be explained by
the presence of the dark radiation and dark pressure was
considered in \citet{Pal05}.

It is the purpose of this paper to consider the geometric
properties of the space-time at the galactic level on the brane,
and to derive the expressions of the physical quantities (dark
radiation and dark pressure), which determine the dynamic of the
particle in circular orbit. Under the assumption of spherical
symmetry the basic equations describing the static gravitational
field on the brane depend on two unknown parameters, the dark
radiation and the dark pressure. As a starting point in our study
we adopt the well-established observation of the constancy of the
galactic rotation curves far-away from the galactic center. This
property allows the immediate determination of the exact galactic
metric on the brane and, consequently, of the mathematical form of
the dark radiation and dark pressure terms. The behavior of other
physical parameters (effective potential, angular momentum and
energy) is also considered.

As a direct observational application of the obtained results we
consider the bending of light and the lensing by galaxies in the
flat rotation curves region. It has already been pointed out that
braneworld black holes could have significantly different
gravitational lensing observational signatures as compared to the
Schwarzschild black holes \citep{Wh04}. A general expression for
the bending angle in the flat rotation curves region is derived
and the deflection of light is studied numerically. The deflection
angle depends on the tangential velocity of the particles in
stable circular orbits, the baryonic mass and the radius of the
galaxy. An analytic expression for the deflection angle in the
first order of approximation is also obtained. The size of the
radius for which the effects of the extra-dimensions are important
is also derived as a function of the tangential velocity and of
the cosmological parameters. The explicit expressions of
observationally important parameters, like the tangential shear
are presented. Hence the theoretical predictions of the deflection
of light in the brane world models can be compared with the
observations.

The predictions of the brane world model are compared with the
observational data in two cases. First, we compare the
observational values of the truncation parameter (the size of the
dark matter halos), obtained in the framework of the truncated
isothermal sphere model by \citet{Ho04} with the size of the
region for which bulk effect are important on the brane. There is
a relatively good agreement between these two quantities, the
difference between prediction and observation being in the range
of $15-20\%$. Secondly, we compare the tangential shear in the
present model with the observational values of \citet{Ho04}.
Consistency between brane world models and observations is
obtained if the mass-radius ratio of the galactic baryonic matter
is of the order of $10^{-4.5}-10^{-4}$. Therefore the bending of
light could provide a powerful method to distinguish between
models which assume that dark matter is a form of unknown matter
or is a result of a change in the dynamical laws that govern the
motion of particles.

The introduction in the past few years of new observational
techniques moved cosmology into the era of precision science. From
the study of the cosmic microwave background (CMB), large scale
structure of galaxies (LSS) and distant type Ia supernovae, a new
paradigm of cosmology has been established. In this new standard
model, the geometry of the Universe is flat so that $\Omega _{\rm
total}=1$, and the total density is made up of matter, (comprised
of baryons ($\Omega _b=0.005$ and cold dark matter ($\Omega
_{CDM}=0.25$) and dark energy with $\Omega _{\Lambda }=0.7$
\citep{Teg04}. In particular, current cosmological data provides a
very precise bound on the physical dark matter density, $\Omega
_{\rm dm}h^2=0.115\pm 0.012$ \citep{Teg04}. This bound provides a
very strong input on any particle physics or alternative
gravitation model for dark matter. On the other hand, dark matter
is assumed to play a very important role in galaxy and large scale
structure formation. Since in the model developed in the present
paper the basic properties of the galactic rotation curves are
explained without resorting to dark matter, it seems that this
would imply that all the dark matter in the Universe could be
related to extra-dimensional or purely non-standard gravitational
effects, as it has been already proposed \citep{Cem03,Marot04}. In
these models the branons, hypothetical particles attached to the
scalar fluctuations of the brane, are decoupled from standard
model matter, thus playing the role of the dark matter. Therefore,
a particle component of the cosmological dark matter cannot be
excluded {\it a priori} even in the framework of brane world
models. Moreover, non-baryonic particles from the standard model
or some of its extensions could also contribute substantially to
the cosmological dynamics.

The present paper is organized as follows. The gravitational field
equations for a static, spherically symmetric vacuum brane are
written down in Section II. The definition and main properties of
the observationally important parameters for the study of the
galactic rotation curves are presented in Section III. In Section
IV we derive the galactic metric and the basic physical parameters
(dark radiation, dark pressure, effective potential etc.) on the
brane. The deflection of light in the flat rotation curves region
is considered in Section V. In Section VI we discuss and conclude
our results.

\section{The field equations for a static, spherically symmetric vacuum brane}

We start by considering a $5$-dimensional space-time (the bulk),
with a single $4$-dimensional brane, on which gravity is confined.
The $4-$D brane
world $\left( ^{(4)}M,g_{\mu \nu }\right) $ is located at a hypersurface $%
\left( B\left( X^{A}\right) =0\right) $ in the $5-$D bulk
space-time $\left(
^{(5)}M,g_{AB}\right) $, of which coordinates are described by $%
X^{A},A=0,1,...,4$. The action of the system is given by
\citep{SMS00}
\begin{equation}
S=S_{bulk}+S_{brane},
\end{equation}
where
\begin{equation}
S_{bulk}=\int_{^{(5)}M}\sqrt{-^{(5)}g}\left[ \frac{1}{2k_{5}^{2}}%
^{(5)}R+^{(5)}L_{m}+\Lambda _{5}\right] d^{5}X,
\end{equation}
and
\begin{equation}
S_{brane}=\int_{^{(4)}M}\sqrt{-^{(5)}g}\left[
\frac{1}{k_{5}^{2}}K^{\pm }+L_{brane}\left( g_{\alpha \beta },\psi
\right) +\lambda _{b}\right] d^{4}x,
\end{equation}
where $k_{5}^{2}=8\pi G_{5}$ is the $5$-D gravitational constant,
$^{(5)}R$ and $^{(5)}L_{m}$ are the $5-$D scalar curvature and
matter Lagrangian in the bulk, respectively. $x^{\mu },\mu
=0,1,2,3$ are the induced $4-$D coordinates of the brane, $K^{\pm
}$ is the trace of the extrinsic curvature on either side of the
brane and $L_{brane}\left( g_{\alpha \beta },\psi \right) $ is the
$4-$D Lagrangian, which is given by a generic functional of the
brane metric $g_{\alpha \beta }$ and of the matter fields $\psi $.
In the following capital Latin indices run in the range $0,...,4$,
while Greek indices take the values $0,...,3$. $\Lambda _{5}$ and
$\Lambda $ are the negative vacuum energy densities in the bulk
and in the brane, respectively.

The Einstein field equations in the bulk are given by
\citep{SMS00,Sa00,Ma03}
\begin{equation}
^{(5)}G_{IJ}=k_{5}^{2}{\rm }^{(5)}T_{IJ},\qquad ^{(5)}T_{IJ}=-\Lambda _{5}%
{\rm }^{(5)}g_{IJ}+\delta (B)\left[ -\lambda _{b}{\rm }%
^{(5)}g_{IJ}+T_{IJ}\right] ,
\end{equation}
where
\begin{equation}
^{(5)}T_{IJ}\equiv -2\frac{\delta ^{(5)}L_{m}}{\delta ^{(5)}g^{IJ}}%
+^{(5)}g_{IJ}{\rm }^{(5)}L_{m},
\end{equation}
is the energy-momentum tensor of bulk matter fields, while $T_{\mu
\nu }$ is the energy-momentum tensor localized on the brane and
which is defined by
\begin{equation}
T_{\mu \nu }\equiv -2\frac{\delta L_{brane}}{\delta g^{\mu \nu
}}+g_{\mu \nu }{\rm }L_{brane}.
\end{equation}

The delta function $\delta \left( B\right) $ denotes the
localization of brane contribution. In the $5$-D space-time a
brane is a fixed point of the $Z_{2}$ symmetry. The basic
equations on the brane are obtained by projection of the variables
onto the brane world. The induced $4-$D metric is $%
g_{IJ}=^{(5)}g_{IJ}-n_{I}n_{J}$, where $n_{I}$ is the space-like
unit vector field normal to the brane hypersurface $^{(4)}M$. In
the following we assume that all the matter fields, except
gravitation, are confined to the brane. This implies that
$^{(5)}L_{m}=0$.

Assuming a metric of the form
$ds^{2}=(n_{I}n_{J}+g_{IJ})dx^{I}dx^{J}$, with $n_{I}dx^{I}=d\chi
$ the unit normal to the $\chi =$constant hypersurfaces and
$g_{IJ}$ the induced metric on $\chi =$constant hypersurfaces, the
effective four-dimensional gravitational equations on the brane
(the Gauss equation), take the form:
\begin{equation}
G_{\mu \nu }=-\Lambda g_{\mu \nu }+k_{4}^{2}T_{\mu \nu
}+k_{5}^{4}S_{\mu \nu }-E_{\mu \nu },  \label{Ein}
\end{equation}
where $S_{\mu \nu }$ is the local quadratic energy-momentum
correction
\begin{equation}
S_{\mu \nu }=\frac{1}{12}TT_{\mu \nu }-\frac{1}{4}T_{\mu
}{}^{\alpha }T_{\nu \alpha }+\frac{1}{24}g_{\mu \nu }\left(
3T^{\alpha \beta }T_{\alpha \beta }-T^{2}\right) ,
\end{equation}
and $E_{\mu \nu }$ is the non-local effect from the free bulk
gravitational
field, the transmitted projection of the bulk Weyl tensor $C_{IAJB}$, $%
E_{IJ}=C_{IAJB}n^{A}n^{B}$, with the property $E_{IJ}\rightarrow
E_{\mu \nu }\delta _{I}^{\mu }\delta _{J}^{\nu }\quad $as$\quad
\chi \rightarrow 0$ \citep{SMS00,Sa00,Ma03}. We have also denoted
$k_{4}^{2}=8\pi G$, with $G$ the usual four-dimensional
gravitational constant.

The four-dimensional cosmological constant, $\Lambda $, and the
four-dimensional gravitational coupling constant, $k_{4}$, are
given by $\Lambda
=k_{5}^{2}\left( \Lambda _{5}+k_{5}^{2}\lambda _{b}^{2}/6\right) /2$ and $%
k_{4}^{2}=k_{5}^{4}\lambda _{b}/6$, respectively. In the limit
$\lambda _{b}^{-1}\rightarrow 0$ we recover standard general
relativity.

The Einstein equation in the bulk and the Codazzi equation also
imply the
conservation of the energy-momentum tensor of the matter on the brane, $%
D_{\nu }T_{\mu }{}^{\nu }=0$, where $D_{\nu }$ denotes the brane
covariant derivative. Moreover, from the contracted Bianchi
identities on the brane it follows that the projected Weyl tensor
should obey the constraint $D_{\nu }E_{\mu }{}^{\nu
}=k_{5}^{4}D_{\nu }S_{\mu }{}^{\nu }$.

The symmetry properties of $E_{\mu \nu }$ imply that in general we
can decompose it irreducibly with respect to a chosen $4$-velocity
field $u^{\mu }$ as \citep{Mar04}
\begin{equation}
E_{\mu \nu }=-{\tilde k}^{4}\left[ U\left( u_{\mu }u_{\nu
}+\frac{1}{3}h_{\mu \nu }\right) +P_{\mu \nu }+2Q_{(\mu }u_{\nu
)}\right] ,  \label{WT}
\end{equation}
where ${\tilde k}=k_{5}/k_{4}$, $h_{\mu \nu }=g_{\mu \nu }+u_{\mu
}u_{\nu }$ projects orthogonal to $u^{\mu }$, the ''dark
radiation'' term $U=-{\tilde k}^{4}E_{\mu \nu }u^{\mu }u^{\nu }$
is a scalar, $Q_{\mu }=k_4^{4}h_{\mu }^{\alpha }E_{\alpha
\beta }$ a spatial vector and $P_{\mu \nu }=-{\tilde k}^{4}\left[ h_{(\mu }{\rm }%
^{\alpha }h_{\nu )}{\rm }^{\beta }-\frac{1}{3}h_{\mu \nu }h^{\alpha \beta }%
\right] E_{\alpha \beta }$ a spatial, symmetric and trace-free
tensor.

In the case of the vacuum state we have $\rho =p=0$, $T_{\mu \nu
}\equiv 0$ and consequently $S_{\mu \nu }=0$. Therefore the field
equations describing a static brane take the form
\begin{equation}
R_{\mu \nu }=\Lambda g_{\mu \nu }-E_{\mu \nu },
\end{equation}
with the trace $R$ of the Ricci tensor $R_{\mu \nu }$ satisfying
the condition $R=R_{\mu }^{\mu }=4\Lambda $. $E_{\mu \nu }$ is a
traceless tensor, $E_{\mu }^{\mu }=0$.

In the vacuum case $E_{\mu \nu }$ satisfies the constraint $D_{\nu
}E_{\mu
}{}^{\nu }=0$. In an inertial frame at any point on the brane we have $%
u^{\mu }=\delta _{0}^{\mu }$ and $h_{\mu \nu }=$diag$\left( 0,1,1,1\right) $%
. In a static vacuum $Q_{\mu }=0$ and the constraint for $E_{\mu
\nu }$ takes the form \citep{GeMa01}
\begin{equation}
\frac{1}{3}D_{\mu }U+\frac{4}{3}UA_{\mu }+D^{\nu }P_{\mu \nu
}+A^{\nu }P_{\mu \nu }=0,
\end{equation}
where $D_{\mu }$ is the projection (orthogonal to $u^{\mu }$) of
the covariant derivative and $A_{\mu }=u^{\nu }D_{\nu }u_{\mu }$
is the 4-acceleration.

In the static spherically symmetric case we may chose $%
A_{\mu }=A(r)r_{\mu }$ and $P_{\mu \nu }=P(r)\left( r_{\mu }r_{\nu }-\frac{1%
}{3}h_{\mu \nu }\right) $, where $A(r)$ and $P(r)$ (the ''dark
pressure'') are some scalar functions of the radial distance $r$,
and $r_{\mu }$ is a unit radial vector \citep{Da00}.

We choose the static spherically symmetric metric on the brane in
the standard form
\begin{equation}
ds^{2}=-e^{\nu \left( r\right) }dt^{2}+e^{\lambda \left( r\right)
}dr^{2}+r^{2}\left( d\theta ^{2}+\sin ^{2}\theta d\phi ^{2}\right)
. \label{line}
\end{equation}

Then the gravitational field equations and the effective
energy-momentum tensor conservation equation in the vacuum take
the form \citep{Ha03,Ma04,Ha04}
\begin{equation}
-e^{-\lambda }\left( \frac{1}{r^{2}}-\frac{\lambda ^{\prime }}{r}\right) +%
\frac{1}{r^{2}}=\frac{48\pi G}{k_4^{4}\lambda _{b}}U+\Lambda,
\label{f1}
\end{equation}
\begin{equation}
e^{-\lambda }\left( \frac{\nu ^{\prime }}{r}+\frac{1}{r^{2}}\right) -\frac{1%
}{r^{2}}=\frac{16\pi G}{k_4^{4}\lambda _{b}}\left( U+2P\right)
-\Lambda, \label{f2}
\end{equation}
\begin{equation}
e^{-\lambda }\left( \nu ^{\prime \prime }+\frac{\nu ^{\prime 2}}{2}+\frac{%
\nu ^{\prime }-\lambda ^{\prime }}{r}-\frac{\nu ^{\prime }\lambda ^{\prime }%
}{2}\right) =\frac{32\pi G}{k_4^{4}\lambda _{b}}\left(
U-P\right)-2\Lambda  , \label{f3}
\end{equation}
\begin{equation}
\nu ^{\prime }=-\frac{U^{\prime }+2P^{\prime
}}{2U+P}-\frac{6P}{r\left( 2U+P\right) },  \label{f4}
\end{equation}
where we denoted $^{\prime }=d/dr$.

Eq. (\ref{f1}) can immediately be integrated to give
\begin{equation}
e^{-\lambda }=1-\frac{C_{1}}{r}-\frac{2GM_U\left( r\right)
}{r}-\frac{\Lambda }{3}r^2, \label{m1}
\end{equation}
where $C_{1}$ is an arbitrary constant of integration, and we
denoted
\begin{equation}\label{mu}
M_U\left( r\right) =\frac{24\pi }{k_4^{4}\lambda _b}\int _0 ^r
r^{2}U\left( r\right) dr.
\end{equation}

The function $M_U(r)$ is the gravitational mass corresponding to
the dark radiation term (the dark mass). For $U=0$ and $\Lambda
=0$ the metric coefficient given by Eq. (\ref{m1}) must tend to
the standard general relativistic Schwarzschild metric
coefficient, which gives $C_{1}=2GM$, where $M=$ constant is the
mass of the gravitating body.

\section{Stable circular orbits and frequency shifts in static space-times on the brane
}

The galactic rotation curves provide the most direct method of
analyzing the gravitational field inside a spiral galaxy. The
rotation curves have been determined for a great number of spiral
galaxies. They are obtained by measuring the frequency shifts $z$
of the light emitted from stars and from the 21-cm radiation
emission from the neutral gas clouds. Usually the astronomers
report the resulting $z$ in terms of a velocity field $v_{tg}$
\citep{Bin87,Pe96,Bo01}.

The starting point in the analysis of the motion of the stars on
the brane is to assume, as usual, that stars behave like test
particles, which follow geodesics of a static and spherically
symmetric space-time. Next, we consider two observers $O_{E{\rm
}}$ and $O_{\infty }$, with four-velocities $u_{E}^{\mu }$ and
$u_{\infty }^{\mu }$, respectively. Observer $O_{E{\rm }}$
corresponds to the light emitter (i. e., to the stars placed at a
point $P_{E}$ of the space-time on the brane), and $O_{\infty}$
represent the detector at point $P_{\infty }$ located far from the
emitter and that can be idealized to correspond to ''spatial
infinity''.

Without loss of generality, we can assume that the stars move in
the galactic plane $\theta =\pi /2$, so that $u_{E}^{\mu }=\left( \dot{t},\dot{r}%
,0,\dot{\phi}\right) _{E}$, where the dot stands for derivation
with respect to the affine parameter $\tau $ along the geodesics. In the timelike case $%
\tau $ corresponds to the proper time. On the other hand, we
suppose that the detector is static (i.e., $O_{\infty }$'s
four-velocity is tangent to the static Killing field $\partial
/\partial t$), and in the chosen coordinate system its
four-velocity is $u_{\infty }^{\mu }=\left( \dot{t},0,0,0\right)
_{\infty }$ \citep{Nuc01}.

The motion of a test particle in the gravitational field on the
brane can be described by the Lagrangian \citep{Ma04}
\begin{equation}
2L=\left( \frac{ds}{d\tau }\right) ^{2}=-e^{\nu \left( r\right)
}\left(
\frac{dt}{d\tau }\right) ^{2}+e^{\lambda \left( r\right) }\left( \frac{dr}{%
d\tau }\right) ^{2}+r^{2}\left( \frac{d\Omega }{d\tau }\right)
^{2},
\end{equation}
where $d\Omega ^{2}=d\theta ^{2}+\sin ^{2}\theta d\phi ^{2}$. From
the Lagrange equations it follows that we have two constants of
motion, the energy $E=e^{\nu (r)}\dot{t}$ and the angular momentum
$l=r^{2}\dot{\phi}$ \citep{La03}.
The condition $u^{\mu }u_{\mu }=-1$ gives $-1=-e^{\nu \left( r\right) }\dot{t%
}^{2}+e^{\lambda (r)}\dot{r}^{2}+r^{2}\dot{\phi}^{2}$ and, with
the use of the constants of motion we obtain
\begin{equation}\label{energy}
E^{2}=e^{\nu +\lambda }\dot{r}^{2}+e^{\nu }\left( \frac{l^{2}}{r^{2}}%
+1\right) .
\end{equation}

This equation shows that the radial motion of the particles on a
geodesic is the same as that of a particle with position dependent mass and with energy $%
E^{2}/2$ in ordinary Newtonian mechanics moving in the effective potential $%
V_{eff}\left( r\right) =e^{\nu (r)}\left( l^{2}/r^{2}+1\right) $.
The conditions for circular orbits $\partial V_{eff}/\partial r=0$
and $\dot{r}=0 $ lead to \citep{La03}
\begin{equation}\label{cons}
l^{2}=\frac{1}{2}\frac{r^{3}\nu ^{\prime }}{1-\frac{r\nu ^{\prime }}{2}}%
,E^{2}=\frac{e^{\nu }}{1-\frac{r\nu ^{\prime }}{2}}.
\end{equation}

The rotation curves of spiral galaxies are inferred from the red
and blue shifts of the emitted radiation by stars moving in
circular orbits on both sides of the central region. The light
signal travels on null geodesics with tangent $k^{\mu }$. We may
restrict $k^{\mu }$ to lie in the equatorial plane $\theta =\pi
/2$ and evaluate the frequency shift for a light signal emitted
from $O_{E}$ in circular orbit and detected by $O_{\infty }$. The
frequency shift associated to the emission and detection of the
light signal is given by
\begin{equation}
z=1-\frac{\omega _{E}}{\omega _{\infty }},
\end{equation}
where $\omega _{I}=-k_{\mu }u_{I}^{\mu }$, and the index $I$
refers to emission ($I=E$) or detection ($I=\infty $) at the
corresponding space-time point \citep{Nuc01,La03}. Two frequency
shifts corresponding to maximum and minimum values are associated
with light propagation in the same and opposite direction of
motion of the emitter, respectively. Such shifts are frequency
shifts of a receding or approaching star, respectively.
Using the constancy along the geodesic of the product of the Killing field $%
\partial /\partial t$ with a geodesic tangent gives the expressions of the
two shifts as \citep{Nuc01,La03}
\begin{equation}
z_{\pm }=1-e^{\left[\nu _{\infty }-\nu \left( r\right)
\right]/2}\frac{1\mp \sqrt{r\nu ^{\prime }/2}}{\sqrt{1-r\nu
^{\prime }/2}},
\end{equation}
respectively, where $\exp \left[ \nu \left( r\right) \right] $
represents the value of the metric potential at the radius of emission $r$ and $\exp %
\left[ \nu _{\infty }\right] $ represents the corresponding value
of $\exp \left[ \nu \left( r\right) \right] $ for $r\rightarrow
\infty $. It is convenient
to define two other quantities $z_{D}=\left( z_{+}-z_{-}\right) /2$ and $%
z_{A}=\left( z_{+}+z_{-}\right) /2$, given by
\begin{equation}\label{zd}
z_{D}\left( r\right) =e^{\left[\nu _{\infty }-\nu \left( r\right)
\right]/2}\frac{\sqrt{r\nu ^{\prime }/2}}{\sqrt{1-r\nu ^{\prime
}/2}},
\end{equation}
\begin{equation}
z_{A}\left( r\right) =1-\frac{e^{\left[\nu _{\infty }-\nu \left( r\right)\right]/2}}{\sqrt{%
1-r\nu ^{\prime }/2}},
\end{equation}
respectively, which can be easily connected to the observations
\citep{Nuc01}.
$z_{A}$ and $z_{D}$ satisfy the relation $\left( z_{A}-1\right) ^{2}-z_{D}^{2}=\exp %
\left[ 2\left( \nu _{D}-\nu \left( r\right) \right) \right] $, and
thus in principle $\exp \left[ \nu \left( r\right) \right] $ can
be obtained directly from the observations. This could provide a
direct observational test of the galactic geometry, and
implicitly, of the brane world and other extra-dimensional models.

The line element on the brane, given by Eq. (\ref{line}), can be
rewritten in terms of the spatial components of the velocity,
normalized with the speed of light, measured by an inertial observer far from the source, as $%
ds^{2}=-dt^{2}\left( 1-v^{2}\right) $ \citep{Ma04}, where
\begin{equation}
v^{2}=e^{-\nu }\left[ e^{\lambda }\left( \frac{dr}{dt}\right)
^{2}+r^{2}\left( \frac{d\Omega }{dt}\right) ^{2}\right] .
\end{equation}

For a stable circular orbit $\dot{r}=0$, and the tangential
velocity of the test particle can be expressed as
\begin{equation}
v_{tg}^{2}=\frac{r^{2}}{e^{\nu }}\left( \frac{d\Omega }{dt}\right)
^{2}.
\end{equation}

In terms of the conserved quantities the angular velocity is given
by
\begin{equation}
v_{tg}^{2}=\frac{e^{\nu }}{r^{2}}\frac{l^{2}}{E^{2}}.
\end{equation}

With the use of Eqs. (\ref{cons}) we obtain
\begin{equation}
v_{tg}^{2}=\frac{r\nu ^{\prime }}{2}.
\end{equation}

Thus, the rotational velocity of the test body is determined by
the metric coefficient $\exp \left( \nu \right) $ only.

\section{Galactic metric, dark radiation and dark pressure on the
brane}

The tangential velocity $v_{tg}$ of stars moving like test
particles around the center of a galaxy is not directly
measurable, but can be inferred from
the redshift $z_{\infty }$ observed at spatial infinity, for which $%
1+z_{\infty }=\exp \left[ \left( \nu _{\infty }-\nu \right)
/2\right] \left( 1\pm v_{tg}\right) /\sqrt{1-v_{tg}^{2}}$. Because
of their non-relativistic velocities in galaxies bounded by
$v_{tg}\leq \left( 4/3\right) \times 10^{-3}$, we observe
$v_{tg}\approx z_{\infty }$ (as the first part of a geometric
series), with the consequence that the lapse $\exp \left( \nu
\right) $ function necessarily tends to unity, i. e., $e^{\nu
}\approx e^{\nu _{\infty }}/\left( 1-v_{tg}^{2}\right) \rightarrow
1$. The
observations show that at distances large enough from the galactic center $%
v_{tg}\approx $ constant \citep{Bin87,Pe96,Bo01}.

In the following we use this observational constraint to
reconstruct the metric of a galaxy on the brane. The condition
\begin{equation}
v_{tg}^{2}=\frac{1}{2}r\nu ^{\prime }={\rm constant},
\end{equation}
immediately allows to find the metric tensor component $e^{\nu }$
in the flat rotation curves region on the brane as
\begin{equation}\label{nu}
e^{\nu }=\left( \frac{r}{R_{b}}\right) ^{2v_{tg}^{2}},
\end{equation}
where $R_{b}$ is a constant of integration.

By adding Eqs. (\ref{f2}) and (\ref{f3}) and eliminating the dark
radiation term $U$ from the resulting equation and Eq. (\ref{f1})
gives the following equation satisfied by the unknown metric
tensor components on the brane:
\begin{equation}
e^{-\lambda }\left( \nu ^{\prime \prime }+\frac{\nu ^{\prime 2}}{2}+\frac{%
2\nu ^{\prime }}{r}-\frac{2\lambda ^{\prime }}{r}-\frac{\nu
^{\prime }\lambda ^{\prime }}{2}+\frac{2}{r^{2}}\right)
-\frac{2}{r^{2}}+4\Lambda =0.
\end{equation}

Substituting $\nu $ given by Eq. (\ref{nu}) leads to a first order
linear differential equation satisfied by $e^{-\lambda }$ and
which is given by
\begin{equation}
\frac{d}{dr}e^{-\lambda }=-2\frac{v_{tg}^{2}\left( v_{tg}^{2}+1\right) +1}{%
v_{tg}^{2}+2}\frac{1}{r}e^{-\lambda }+\frac{2}{v_{tg}^{2}+2}\frac{1}{r}-\frac{4\Lambda }{%
v_{tg}^{2}+2}r.  \label{lambda}
\end{equation}

From Eq. (\ref{lambda}) we obtain  $e^{-\lambda }$ as
\begin{equation}\label{lam}
e^{-\lambda }=\frac{1}{\left( v_{tg}^{2}+2\right) \alpha
}+C_{2}r^{-2\alpha }-\frac{2\Lambda }{\left( v_{tg}^{2}+2\right)
\left( \alpha +1\right) }r^{2},
\end{equation}
where $C_{2}$ is an arbitrary constant of integration and we
denoted
\begin{equation}
\alpha =\frac{v_{tg}^{2}\left( v_{tg}^{2}+1\right)
+1}{v_{tg}^{2}+2}.
\end{equation}

Eqs. (\ref{nu}) and (\ref{lam}) give the complete galactic metric
on the brane.

Once the metric tensor components are known, the calculation of
the dark radiation and dark pressure terms is straightforward. Eq.
(\ref{f1}) immediately gives the dark radiation $U$ on the brane
as
\begin{equation}
\frac{48\pi G}{k_4^{4}\lambda _{b}}U(r)=\frac{1}{v_{tg}^{2}+2}\left[ \frac{%
2\left( 2-\alpha \right) }{\left( \alpha +1\right)
}-v_{tg}^{2}\right]
\Lambda +\left[ 1-\frac{1}{\left( v_{tg}^{2}+2\right) \alpha }\right] \frac{1%
}{r^{2}}+\left( 2\alpha -1\right) C_{2}r^{-2\alpha -2}.
\end{equation}

From Eqs. (\ref{f1}) and (\ref{f2}) we obtain the dark pressure on
the brane as
\begin{equation}
\frac{96\pi G}{k_4^{4}\lambda _{b}}P(r)=e^{-\lambda }\left(
\frac{3\nu
^{\prime }}{r}+\frac{4}{r^{2}}-\frac{\lambda ^{\prime }}{r}\right) -\frac{4}{%
r^{2}}+4\Lambda ,
\end{equation}
or
\begin{eqnarray}
\frac{96\pi G}{k_4^{4}\lambda _{b}}P(r)
&=&\frac{4}{v_{tg}^{2}+2}\left[ v_{tg}^{2}+1-\frac{2\left(
v_{tg}^{2}+2\right) \alpha +5v_{tg}^{2}+1}{\left(
v_{tg}^{2}+2\right) \left( \alpha +1\right) }\right] \Lambda +\frac{2}{%
v_{tg}^{2}+2}\left[ \frac{5v_{tg}^{2}+1}{\left( v_{tg}^{2}+2\right) \alpha }%
-2v_{tg}^{2}-1\right] \frac{1}{r^{2}} \nonumber\\
&&+2C_{2}\frac{2\left( v_{tg}^{2}+2\right) \alpha +5v_{tg}^{2}+1}{%
v_{tg}^{2}+2}r^{-2\alpha -2}.\
\end{eqnarray}

The angular momentum and the energy of a star moving in the
galactic gravitational field on the brane are given by
\begin{equation}
l=r\frac{v_{tg}}{\sqrt{1-v_{tg}^{2}}},
\end{equation}
and
\begin{equation}
E=\left( \frac{r}{R_{b}}\right)
^{v_{tg}^{2}}\frac{1}{\sqrt{1-v_{tg}^{2}}},
\end{equation}
respectively. For the effective potential describing the Newtonian
motion of a test particle we find
\begin{equation}
V_{eff}\left( r\right) =\left( \frac{r}{R_{b}}\right) ^{2v_{tg}^{2}}\frac{1}{%
1-v_{tg}^{2}}.
\end{equation}

From Eq. (\ref{mu}) it follows that the dark mass associated to
the dark radiation term is given by
\begin{equation}
2GM_{U}\left( r\right) =\frac{1}{3\left( v_{tg}^{2}+2\right) }\left[ \frac{%
2\left( 2-\alpha \right) }{\left( \alpha +1\right)
}-v_{tg}^{2}\right]
\Lambda r^{3}+\left[ 1-\frac{1}{\left( v_{tg}^{2}+2\right) \alpha }\right] r-%
C_{2}r^{-2\alpha -1}.  \label{massu}
\end{equation}

At distances relatively close to the galactic center, where the
effect of the cosmological constant can be neglected, but
sufficiently high so that the last term in Eq. (\ref{massu})  is
also negligible small, the dark mass can be approximated as
\begin{equation}\label{m2}
2GM_{U}\left( r\right) \approx \frac{v_{tg}^{2}\left( v_{tg}^{2}+1\right) }{%
v_{tg}^{2}+2}r\approx \frac{v_{tg}^{2}}{2}r.
\end{equation}

The dark mass is linearly increasing with the radial distance from
the galactic center.

The behavior of the metric coefficients and of the dark radiation
and pressure in the solutions we have obtained depends on two
arbitrary constants of integration $R_b$ and $C_2$. Their
numerical value can be obtained by assuming the continuity of the
metric coefficient $\exp \left( \lambda \right) $ across the
vacuum boundary of the galaxy. For simplicity we assume that
inside the "normal" (baryonic) luminous matter, with density $\rho
_B$, which form a galaxy, the non-local effects of the Weyl tensor
can be neglected. We define the vacuum boundary $r_0$ of the
galaxy (which for simplicity is assumed to have spherical
symmetry) by the condition $\rho _B \left(r_0\right)\approx 0$.

Therefore at the vacuum boundary of the galaxy the metric
coefficients are $\exp \left( \nu \right)=1-2GM_B/r_0$ and $\exp
\left( -\lambda \right)=1-2GM_B/r_0$, where $M_{B}=4\pi
\int_{0}^{r_{0}}\rho _{B}\left( r\right) r^{2}dr$ is the total
baryonic mass inside the radius $r_0$. The continuity of $\exp
\left( \nu \right) $ through the surface $r=r_0$ gives
\begin{equation}\label{condr}
R_{b}=r_{0}\left( 1-\frac{2GM_{B}}{r_{0}}\right)
^{-\frac{1}{2}v_{tg}^{-2}},
\end{equation}
while the continuity of $\exp \left( -\lambda \right) $ fixes the
integration constant $C_2$ as
\begin{equation}
C_{2}=\frac{r_{0}^{2\alpha }}{v_{tg}^{2}+2}\left[ \left( 1-\frac{2GM_{B}}{%
r_{0}}\right) \left( v_{tg}^{2}+2\right) -\frac{1}{\alpha }+\frac{2\Lambda }{%
\alpha +1}r_{0}^{2}\right] .
\end{equation}

Thus at the galactic level the metric coefficients and the dark
radiation and pressure on the brane can be obtained in terms of
observable quantities.

\section{Light deflection and lensing by galaxies in brane world models}

One of the ways we could in principle test the galactic metric
obtained in the previous Section would be by studying the light
deflection by galaxies, and in particular by studying the
deflection of photons passing through the region where the
rotation curves are flat. Let us consider a photon approaching a
galaxy from far distances. We will compute the deflection by
assuming that the metric is given by Eqs. (\ref{nu}) and
(\ref{lam}).

The bending of light on the brane by the galactic gravitational
field results in a deflection angle $\Delta \phi $ given by
\begin{equation}\label{defl1}
\left(\Delta \phi\right)_{brane} =2\left| \phi \left( r_{0}\right)
-\phi _{\infty }\right| -\pi ,
\end{equation}
where $\phi _{\infty }$ is the incident direction and $r_{0}$ is
the coordinate radius of the closest approach to the center of the
galaxy, and generally \citep{Nuc01}
\begin{equation}
\phi \left( r_{0}\right) -\phi _{\infty }=\int_{r_{0}}^{\infty }e^{\frac{%
\lambda (r)}{2}}\left[ e^{\nu \left( r_{0}\right) -\nu \left(
r\right) }\left( \frac{r}{r_{0}}\right) ^{2}-1\right]
^{-1/2}\frac{dr}{r}.
\end{equation}

By taking into account the explicit expressions of the metric
tensor components in the flat rotation curves region we obtain
\begin{equation}\label{bend}
\phi \left( r_{0}\right) -\phi _{\infty }=\int_{r_{0}}^{\infty }\frac{dr}{r%
\sqrt{\left[ \frac{1}{\left( v_{tg}^{2}+2\right) \alpha }+C_{2}r^{-2\alpha }-%
\frac{2\Lambda }{\left( v_{tg}^{2}+2\right) \left( \alpha +1\right) }r^{2}%
\right] \left[ e^{\nu \left( r_{0}\right)
}r_{0}^{-2}R_{b}^{2v_{tg}^{2}}r^{2\left( 1-v_{tg}^{2}\right)
}-1\right] }}.
\end{equation}

By introducing a new variable $\eta =r/r_{0}$ and taking into
account the matching condition given by Eq. (\ref{cond}) we can
rewrite Eq. (\ref{bend}) as
\begin{equation}\label{defl11}
\phi \left( r_{0}\right) -\phi _{\infty }=\int_{1}^{\infty }\frac{\sqrt{%
v_{tg}^{2}+2}\eta ^{-1}\left[ \eta ^{2\left( 1-v_{tg}^{2}\right)
}-1\right] ^{-1/2}d\eta }{\sqrt{\left[ \frac{1}{\alpha }+\eta
^{-2\alpha }\left[ \left(
1-\frac{2GM_{B}}{r_{0}}\right) \left( v_{tg}^{2}+2\right) -\frac{1}{\alpha }+%
\frac{2\Lambda }{\alpha +1}r_{0}^{2}\right] -\frac{2\Lambda r_{0}^{2}}{%
\alpha +1}\eta ^{2}\right] }}.
\end{equation}

In standard general relativity the bending angle by a galaxy is given by $%
\left( \Delta \phi \right) _{GR}=2\left| \phi \left( r_{0}\right)
-\phi _{\infty }\right| -\pi =4GM_{B}/r_{0}$. The total deflection
angle is given by $\left( \Delta \phi \right) _{total}=\left(
\Delta \phi \right) _{brane}+\left( \Delta \phi \right) _{GR}$.

In order to estimate the magnitude of the bulk contribution to the
bending of light on the brane and to compare this contribution to
the results from standard dark matter models we define the
parameter
\begin{equation}\label{delta}
\delta =\frac{\left( \Delta \phi \right) _{brane}}{\left( \Delta
\phi \right) _{DM}},
\end{equation}
where $\left( \Delta \phi \right) _{DM}$ is the deflection angle
as is usually considered in the standard dark matter models.
Generally, in the dark halo models, a light ray which goes past
the halo without entering it propagates entirely in a
Schwarzschild metric and the light ray is deflected by an angle
$\left( \Delta \phi \right) _{GR}$. The bending of the light which
passes through the dark matter halo is determined by the metric
inside the halo and this depends on the assumed properties of the
dark matter.

 We shall compute the parameter $\delta $ in a semi-realistic
model for dark matter, in which it is assumed that the galaxy (the
baryonic matter) is embedded into an isothermal mass distribution
(the dark matter),
with the density varying as $\rho =\sigma _{v}^{2}/2\pi Gr^{2}$, where $%
\sigma _{v}$ is the line of sight velocity dispersion
\citep{Bin87}. In this model it is assumed that the mass
distribution of the dark matter is spherically symmetric. In fact,
if the rotation curve is flat, then the mass
distribution must be that of the isothermal sphere, for which we also have $%
v_{tg}=\sqrt{2}\sigma _{v}$. The surface density $\Sigma $ of the
isothermal sphere is $\Sigma \left( r\right) =\sigma
_{v}^{2}/2Gr$. For this dark
matter distribution the bending angle of light is constant and is given by $%
\left( \Delta \phi \right) _{DM}=2\pi v_{tg}^{2}$ \citep{BlNa92}.

Therefore in this model
\begin{equation}
\delta =\frac{\left( \Delta \phi \right) _{brane}}{2\pi
v_{tg}^{2}}.
\end{equation}

The variation of the parameter $\delta $ as a function of the
quantity $\chi =4GM_B/r_0$ is represented in Fig. 1.

In order to find an approximate analytic expression for the
deflection of light due to the presence of the Weyl stresses on
the brane we perform a Taylor series expansion of the function
$F(\eta )=\left( 1/\left( v_{tg}^{2}+2\right) \alpha +C_{2}\eta
^{-2\alpha }-2\Lambda r_{0}^{2}\eta ^{2}/\left(
v_{tg}^{2}+2\right) \left( \alpha +1\right) \right) ^{-1/2}$ near
$\eta =1$. In the first order we obtain
\begin{equation}
F\left( \eta \right) \approx \left[ \frac{1}{\left(
v_{tg}^{2}+2\right) \alpha }+C_{2}r_0^{-2\alpha }-\frac{2\Lambda
r_{0}^{2}}{\left( v_{tg}^{2}+2\right) \left( \alpha +1\right)
}\right] ^{-1/2}-\frac{2\left( C_{2}r_0^{-2\alpha }\alpha
+\frac{2\Lambda r_{0}^{2}}{\left( v_{tg}^{2}+2\right) \left(
\alpha +1\right) }\right)
\left( \eta -1\right) }{\left[ \frac{1}{\left( v_{tg}^{2}+2\right) \alpha }%
+C_{2}r_0^{-2\alpha }-\frac{2\Lambda r_{0}^{2}}{\left(
v_{tg}^{2}+2\right) \left( \alpha +1\right) }\right] ^{3/2}}+....
\end{equation}

Substituting this expression into Eq. (\ref{defl11}) gives, after
evaluating the integral, the first order gravitational deflection
angle of the light on the brane as
\begin{eqnarray}
\left( \Delta \phi \right) _{brane} &=&2\left|\frac{\pi
\sqrt{\frac{1}{\left( v_{tg}^{2}+2\right) \alpha
}+C_{2}-\frac{2\Lambda r_{0}^{2}}{\left( v_{tg}^{2}+2\right)
\left( \alpha +1\right) }}}{2\left( 1-v_{tg}^{2}\right) }\right.%
- \nonumber\\
&&\left.\frac{\sqrt{\pi }\left( C_{2}\alpha +\frac{2\Lambda
r_{0}^{2}}{\left(
v_{tg}^{2}+2\right) \left( \alpha +1\right) }\right) \left[ \frac{\sqrt{\pi }%
}{2\left( 1-v_{tg}^{2}\right) }+\frac{\Gamma \left( -\frac{1}{2}\frac{%
v_{tg}^{2}}{1-v_{tg}^{2}}\right) }{\Gamma \left( -\frac{1}{2}\frac{1}{%
1-v_{tg}^{2}}\right) }\right] }{\left( \frac{1}{\left(
v_{tg}^{2}+2\right) \alpha }+C_{2}-\frac{2\Lambda
r_{0}^{2}}{\left( v_{tg}^{2}+2\right) \left( \alpha +1\right)
}\right) ^{3/2}}\right|-\pi ,
\end{eqnarray}
where $\Gamma (z)$ is the Euler gamma function, defined as $\Gamma
\left( z\right) =\int_{0}^{\infty }t^{z-1}e^{-t}dt$.

Alternatively, some very simple expressions for the parameter
$\delta $ can be derived in the limits $\chi >>v_{tg}^{2}$ and
$\chi <<v_{tg}^{2}$, respectively. By assuming that
$v_{tg}^{2}<<1$ and neglecting all terms containing the tangential
velocity and the cosmological constant, the light deflection angle
on the brane becomes
\begin{equation}
\left( \Delta \phi \right) _{brane}\approx 2\left|
\int_{1}^{\infty }\eta ^{-1}\left[ \left( \eta ^{2}-1\right)
\left( 2-\chi \eta ^{-1}\right) \right] ^{-1/2}d\eta \right| -\pi
.
\end{equation}

The integral can be calculated exactly, and we obtain
\begin{equation}
\left( \Delta \phi \right) _{brane}\approx \left| \frac{\sqrt{2}}{4}\frac{4%
\sqrt{2}K\left( \frac{2\chi }{\chi -2}\right) +_{3}F_{2}\left[ \left\{ \frac{%
3}{4},1,\frac{5}{4}\right\} ,\left\{ \frac{3}{2},\frac{3}{2}\right\} ,\frac{%
\chi ^{2}}{4}\right] \sqrt{2-\chi }\chi }{\left( 1-\frac{\chi
}{2}\right) ^{1/2}}\right| -\pi ,  \label{approx1}
\end{equation}
where $K(m)$ is the complete elliptic integral of the first kind given by $%
K(m)=\int_{0}^{\pi /2}\left( 1-m\sin ^{2}\theta \right)
^{-1/2}d\theta $ and $_{p}F_{q}\left( \vec{a};\vec{b};z\right) $
is the generalized hypergeometric function defined as
\begin{equation}
_{p}F_{q}\left( \vec{a};\vec{b};z\right) =\sum_{k=0}^{\infty
}\left( a_{1}\right) _{k}...\left( a_{p}\right) _{k}/\left(
b_{1}\right) _{k}...\left( b_{q}\right) _{k}z^{k}/k!.
\end{equation}

A series expansion of Eq. (\ref{approx1}) gives
\begin{equation}
\left( \Delta \phi \right) _{brane}\approx \frac{\chi }{2}.
\end{equation}

Therefore for the isothermal dark matter model and in the limit of
large $\chi $ the parameter $\delta $ tends to the value $\chi
/4\pi v_{tg}^2$, $\delta \rightarrow \chi /4\pi v_{tg}^2$.

Neglecting $\chi $ and the cosmological constant in Eq.
(\ref{defl11}) gives
\begin{equation}
\left( \Delta \phi \right) _{brane}\approx 2\left|
\int_{1}^{\infty }\eta
^{-1}\left[ \left( \eta ^{2}-1\right) \left( 2+v_{tg}^{2}\eta ^{-1}\right) %
\right] ^{-1/2}d\eta \right| -\pi .
\end{equation}

A similar calculation as above gives  $\left( \Delta \phi \right)
_{brane}\approx v_{tg}^{2}/2$. Hence in the limit of small $\chi
$, for the isothermal dark matter model $\delta $ can be
approximated as $\delta \approx 1/4\pi $. These two limiting forms
of the parameter $\delta $ are consistent with the numerical
results presented in Fig. 1.

If the effect of the cosmological constant can be neglected, the
deflection angle of the light in the Weyl stresses dominated
region around a galaxy is a function of the tangential velocity of
the test particles in stable circular orbit only. To obtain the
total value of the bending angle of the light one must also add
the usual general relativistic contribution to the bending from
the baryonic mass of the galaxy.

Once the light deflection angle is known, one can study the
gravitational lensing on the brane in the flat rotation curves
region. The lensing geometry is illustrated in Fig. 2.

The light emitted by the source $S$ is deflected by the lens $L$
(a galaxy in our case) and reaches the observer $O$ at an angle
$\theta $ to the optic axis $OL$, instead of $\beta $. The lens
$L$ is located at a distance $D_L$ to the observer and a distance
$D_{LS}$ to the source, respectively, while the observer-source
distance is $D_S$. $r_0$ is the impact factor (distance of closest
approach) of the photon beam.

The lens equation is given by \citep{Wh04}
\begin{equation}
\tan \beta =\tan \theta -\frac{D_{LS}}{D_{S}}\left[ \tan \theta
+\tan \left( \Delta \phi -\theta \right) \right] .
\end{equation}

By assuming that the angle $\theta $ is small, we have $\tan
\theta \approx \theta $ and the lens equation can be written as
\begin{equation}
\beta \approx \theta -\frac{D_{LS}}{D_{S}}\Delta \phi .
\end{equation}

In the special case of the perfect alignment of the source, lens
and observers, $\beta =0$, and the azimuthal axial symmetry of the
problem yields a ring image, the Einstein ring, with angular
radius
\begin{equation}
\theta _{E}^{(brane)}\approx \frac{D_{LS}}{D_{S}}\Delta \phi .
\end{equation}

This equation can be expressed in a more familiar form by taking
into account that the impact parameter $r_{0}\approx D_{L}\theta
$, which gives
\begin{equation}
\theta _{E}^{(brane)}\approx \sqrt{\frac{D_{LS}}{D_{S}D_{L}}\Delta
\phi r_{0}}\approx \theta
_E^{(GR)}\sqrt{\frac{r_0}{GM_B}}\sqrt{\Delta \phi },
\end{equation}
where $\theta _E^{(GR)}$ is the angular radius of the Einstein
ring in the case of standard general relativity, $\theta
_{E}^{(GR)}=\sqrt{4\left( D_{LS}/D_{S}D_{L}\right) GM_{B}}$.

In the case of a galaxy with a heavy isothermal dark matter
distribution, the Einstein radius of the lens formed in perfect
alignment is \citep{BlNa92}
\begin{equation}
\theta _{E}^{(DM)}=\left( \frac{4\pi \sigma _{v}^{2}}{c^{2}}\right) \frac{%
D_{LS}}{D_{S}}.
\end{equation}

The ratio $\delta $ of the Einstein's rings angular diameters in
the brane world models and in the isothermal dark galactic halo
model is
\begin{equation}
\delta =\frac{\theta _{E}^{(brane)}}{\theta
_{E}^{(DM)}}=\frac{\left( \Delta \phi \right) _{brane}}{2\pi
v_{tg}^{2}}.
\end{equation}

Therefore the ratio of the angular radii of the Einstein rings in
the brane world models and in the isothermal dark matter model is
given by the same parameter $\delta $ which has been already
introduced in Eq. (\ref{delta}). The variation of the ratio of the
Einstein rings in the two models is presented in Fig. 1.

\section{Discussions and final remarks}

The galactic rotation curves continue to pose a challenge to
present day physics as one would to have a better understanding of
some of the intriguing phenomena associated with them, like their
universality and the very good correlation between the amount of
dark matter and the luminous matter in the galaxy. To explain
these observations models based on particle physics in the
framework of Newtonian gravity are the most commonly considered.

In the present paper we have considered and further developed an
alternative view to the dark matter problem, namely, that the
galactic rotation curves can naturally be explained in models in
which our Universe is a domain wall (a brane) in a
multi-dimensional space-time (the bulk). The extra-terms in the
gravitational field equations on the brane induce a supplementary
gravitational interaction, which can account for the observed
behavior of the galactic rotation curves. By using the simple
observational fact of the constancy of the galactic rotation
curves, the galactic metric and the corresponding Weyl stresses
(dark radiation and dark pressure) can be completely
reconstructed.

The form of the galactic metric we have obtained in the framework
of the brane world models differs from what would be naively expected, that is, $%
ds^{2}=-\left( 1+2\Phi \right) dt^{2}+\left( 1+2\Phi \right)
^{-1}dr^{2}+r^{2}\left( d\theta ^{2}+\sin ^{2}\theta d\phi
^{2}\right) $, with $\Phi $ representing the Newtonian potential.
This form is often implicitly assumed \citep{Pal05}, and the fact
that it is not appropriate for the region where the rotation
curves are flat can lead to significant errors in the estimation
of the magnitude of some important physical effects, like, for
example, the bending of light by the galaxies.

The observations in spiral galaxies usually determine $v_{tg}$
from the redshift $z_D$, so that $z_D\approx v_{tg}\approx $
constant. By assuming that $\exp\left(\nu _{\infty}\right)\approx
1$, and with the use of Eqs. (\ref{zd}) and (\ref{nu}), it follows
that the results we have obtained are self-consistent if the
condition
\begin{equation} \left( \frac{r}{R_{b}}\right)
^{-v_{tg}^{2}}\left( 1-v_{tg}^{2}\right) ^{-1/2}\approx 1,
\label{cond}
\end{equation}
holds. By using the approximation $\left( r/R_{b}\right)
^{-v_{tg}^{2}}\approx 1-v_{tg}^{2}\ln \left( r/R_{b}\right) $, valid for $%
\left| v_{tg}^{2}\ln \left( r/R_{b}\right) \right| <<1$, it
follows that the condition is satisfied if $\left| \ln \left(
r/R_{b}\right) \right| <<v_{tg}^{-2}$. Since $v_{tg}\approx
10^{-3}-10^{-4}$, the approximations we have used are
self-consistent as long as $-10^{6}<<\ln \left( r/R_{b}\right)
<<10^{6}$, condition which, for the case of the galaxies, does not
impose any practically relevant constraint.

As a second consistency condition we require that the timelike
circular geodesics on the brane be stable. Let $r_{0}$ be a
circular orbit and consider a perturbation of it of the form
$r=r_{0}+\delta $, where $\delta
<<r_{0}$ \citep{La03}. Taking expansions of $V_{eff}\left( r\right) $ and $%
\exp \left( \nu +\lambda \right) $ about $r=r_{0}$, it follows
from Eq. (\ref {energy}) that
\begin{equation}
\ddot{\delta}+\frac{1}{2}e^{\nu \left( r_{0}\right) +\lambda
\left( r_{0}\right) }V_{eff}^{\prime \prime }\left( r_{0}\right)
\delta =0.
\end{equation}

The condition for stability of the simple circular orbits requires $%
V_{eff}^{\prime \prime }\left( r_{0}\right) >0$ \citep{La03}. This
gives
\begin{equation}
2\left( \frac{r_{0}}{R_{b}}\right) ^{2v_{tg}^{2}}r_{0}^{-4}\left[
-v_{tg}^{2}\left( 1-2v_{tg}^{2}\right) r_{0}^{2}+l^{2}\left(
3-5v_{tg}^{2}+2v_{tg}^{4}\right) \right] >0.
\end{equation}

By neglecting the small terms containing powers of the tangential velocity $%
v_{tg}$ with respect to the unity, it follows that the circular
orbits are stable if their radius satisfy the condition
$r_{0}^{2}<3l^{2}/v_{tg}^{2}$.

In the present model there is a very simple relation between the
mass of the inter-galactic dark radiation $M_{U}$ and the luminous
(baryonic) mass $M_{B}$ of the galaxy. By assuming that the
tangential velocity of particles in circular orbit is
approximately given by $v_{tg}^{2}\approx GM_{B}/r_{0}$, it
follows that $M_{U}$ is related to $M_{B}$ via the following
simple scaling relation
\begin{equation}
M_{U}(r)\approx \frac{r}{2r_{0}}M_{B}{\rm.}
\end{equation}

The mass of the dark radiation is proportional to the mass of the
galaxy and is linearly increasing with the distance to the
galactic center.

If we assume that the flat rotation curves extend indefinitely,
the resulting space-time is not asymptotically flat, but of de
Sitter type. This is due to the presence of the cosmological
constant $\Lambda $ on the brane. Observationally, the galactic
rotation curves remain flat to the farthest distances that can be
observed. On the other hand there is a simple way to estimate an
upper bound for the cutoff of the constancy of the tangential
velocities. The idea is to consider the point at which the
decaying density profile of the dark radiation associated to the
galaxy becomes smaller than the average energy density of the
Universe. Let the value of the coordinate radius at the point
where the two densities are equal be $R_{U}^{\max }$. Then at this
point $\left(8\pi G/k_4^4\lambda _b\right)U\left( R_{U}^{\max
}\right) =\left(8\pi G/c^2\right)\rho _{univ}$, where $\rho
_{univ}c^2$ is the mean energy density of the universe. Hence we
obtain
\begin{equation}
R_{U}^{\max }=\sqrt{\frac{1}{\frac{8\pi G}{c^{2}}\rho _{univ}-\frac{1}{%
v_{tg}^{2}+2}\left[ \frac{2\left( 2-\alpha \right) }{\left( \alpha
+1\right)
}-v_{tg}^{2}\right] \Lambda }}\frac{\sqrt{v_{tg}^{2}+1}}{\sqrt{%
v_{tg}^{2}\left( v_{tg}^{2}+1\right) +1}}v_{tg}.
\end{equation}

The mean density of the universe and the value of the cosmological
constant can be expressed with the help of the density parameters
$\Omega =\rho _{univ}/\rho _{crit{\rm }}$ and $\Omega _{\Lambda
}=\Lambda c^{2}/3H_{0}^{2} $, respectively, where $\rho
_{crit}=3H_{0}^{2}/8\pi G$, with $H_{0}$ is the Hubble constant,
given by $H_{0}=100h$ km/sec Mpc, $1/2\leq h\leq 1$. Therefore
\begin{equation}\label{ru}
R_{U}^{\max }=\frac{c}{\sqrt{3}}H_{0}^{-1}\sqrt{\frac{1}{\Omega -\frac{1}{%
v_{tg}^{2}+2}\left[ \frac{2\left( 2-\alpha \right) }{\left( \alpha
+1\right)
}-v_{tg}^{2}\right] \Omega _{\Lambda }}}\frac{\sqrt{v_{tg}^{2}+1}}{\sqrt{%
v_{tg}^{2}\left( v_{tg}^{2}+1\right) +1}}v_{tg}.
\end{equation}

A numerical evaluation of $R_{U}^{\max }$ requires the knowledge
of the ratio of the rotation velocities in the flat region and of
the basic fundamental cosmological parameters. For $v_{tg}$ in the
range $v_{tg}\in \left( 10^{-4},10^{-3}\right) $ and for $\Omega
=1$, $\Omega _{\Lambda }=0.7$ we obtain $R_{U}^{\max }\in \left(
315h^{-1},3150h^{-1}\right) $ kpc. On the other hand the measured
flat regions are about $R\approx 2\times R_{opt}$, where $R_{opt}$
is the radius encompassing $83\%$ of the total integrated light of
the galaxy \citep{Bin87,Pe96,Bo01}. If we take as a typical value
$R\approx 30$ kpc, then it follows that $R<R_{U}^{\max }$.
However, according to our model, the flat rotation curves region
should extend far beyond the present measured range.

An alternative estimation of $R_{U}^{\max }$ can be obtained from
the observational requirement that at the cosmological level the
energy density of the dark matter represents a fraction $\Omega
_{m}\approx 0.3$ of the total energy density of the universe
$\Omega =1$. Therefore the dark matter contribution inside a
radius $R_{U}^{\max }$ is given by $4\pi \Omega _{m}\left(
R_{U}^{\max }\right) ^{3}\rho _{crit}/3$, which gives
\begin{equation}
R_{U}^{\max }\approx \sqrt{\frac{1}{2\Omega
_{m}}}\frac{c}{H_{0}}v_{tg}.
\end{equation}

Therefore, by assuming that the dark radiation contribution to the
total energy density of the Universe is of the order of $\Omega
_{m}\approx 0.3$
we have $R_{U}^{\max }\in \left( 388h^{-1},3881h^{-1}\right) $ kpc for $%
v_{tg}\in \left( 10^{-4},10^{-3}\right) $.

The limiting radius at which the effects of the extra-dimensions
extend, far away from the baryonic matter distribution, is given
in the present model by Eq. (\ref{ru}). In the standard dark
matter models this radius is called the truncation parameter $s$,
and it describes the extent of the dark matter halos. Values of
the truncation parameter by weak lensing have been obtained for
several fiducial galaxies by \citet{Ho04}. In the following we
compare our results with the observational values of $s$ obtained
by fitting the observed values with the truncated isothermal
sphere model, as discussed in some detail in \citet{Ho04}. The
truncation parameter $s$ is related to $R_U$ by the relation
$s=R_U/2\pi $ (see Eq. (4) in \citet{Ho04} and Eq. (\ref{m2}) in
the present paper). Therefore, generally $s$ can be obtained from
the relation
\begin{equation}
s\approx \frac{\sigma }{\sqrt{6}\pi }\sqrt{\frac{1}{2\Omega
_{m}}}H_{0}^{-1}, \label{s}
\end{equation}
where $\sigma $ is the velocity dispersion, expressed in km/s.
Hence the truncation parameter is a simple function of the
velocity dispersion and of the cosmological parameters only. For
a velocity dispersion of $\sigma =146$ km/s and with $\Omega _{m}=0.3$, Eq. (%
\ref{s}) gives $s\approx 245h^{-1}$ kpc, while the truncation size
obtained observationally in \citet{Ho04} is $s=213h^{-1}$ kpc. For
$\sigma =110$ km/s we obtain $s\approx 184h^{-1}$ kpc, while
$\sigma =136$ km/s gives $s\approx 228h^{-1}$ kpc. All these
values are consistent with the observational results reported in
\citet{Ho04}, the error between prediction and observation being
of the order of $20\%$. We have also to mention that the
observational values of the truncation parameter depend on the
scaling relation between the velocity dispersion and the fiducial
luminosity of the galaxy. Two cases have been considered in
\citet{Ho04}, the case in which the luminosity $L_B$ does not
evolve with the redshift $z$ and the case in which $L_B$ scales
with $z$ as $L_B\propto (1+z)$. Depending on the scaling relation
slightly different values of the velocity dispersion and
truncation parameter are obtained.

The study of the deflection of light (gravitational lensing) in
the flat rotation curves region can provide a powerful
observational tool for discriminating between standard dark matter
and brane world models. Due to the fixed form of the galactic
metric on the brane, the light bending angle is a function of the
tangential velocity of particles in stable circular orbit and the
baryonic mass and radius of the galaxy. The specific form of the
bending angle is determined by the brane galactic metric, and this
form is very different as compared to the other dark matter models
(long-range self-interacting scalar fields, MOND, non-symmetric
gravity etc.). When $\chi =4GM_B/r_0<<1$, the gravitational light
deflection angle is much larger than the value predicted by the
standard general relativistic approach. Even when we compare our
results with standard dark matter models, like the isothermal dark
matter halo model, we still find significant differences in the
lensing effect. Therefore the study of the gravitational lensing
may provide evidence for the existence of the bulk effects on the
brane. On the other hand, since in this model there is only
baryonic matter, all the physical properties at the galactic level
are determined by the amount of the luminous matter and its
distribution.

Generally, the angular position of the source $\beta $ is related
to the angular position of its image by the lens equation, $\beta
=\theta +\alpha (\theta )$. The standard general relativistic
deflection law for a point mass in general relativity is $\alpha
(\theta )=-\theta _{E}^{2}/\theta $ \cite{MoTu01}. In order to
test alternative theories of gravity a more general point mass
deflection law, of the form $\alpha (\theta )=-\left( \theta
_{E}^{2}/\theta \right) \left( \theta _{0}/\theta +\theta
_{0}\right) ^{\xi -1}$, $\xi \geq 1$, was introduced by
\cite{MoTu01} (see \cite{Ho02} for a more general
parametrization). Here $\theta _{0}$ is a parameter which can be
related to the scale $r_{0}$ beyond which the physics becomes
non-Newtonian.
In the brane world model the function $\alpha $ can be represented as $%
\alpha \left( \theta \right) =-\left[ \left( \theta
_{E}^{(GR)}\right) ^{2}/\theta \right] \left( r_{0}/GM_{B}\right)
\left(\Delta \phi \right)_{brane}$. From the deflection law one
can find the tangential shear of the image, $\gamma _{\tan }\left(
\theta \right) =\left[ \left( d\alpha /d\theta \right) -\alpha
(\theta )/\theta \right] /2$. For the brane world model we obtain
\begin{equation}
\gamma _{\tan }\left( \theta \right)=\frac{\left( \theta
_{E}^{(GR)}\right) ^{2}}{\theta ^{2}}\frac{r_{0}}{GM_{B}}\left(
\Delta \phi \right) _{brane}.
\end{equation}

By using the definition of the parameter $\delta $ we obtain
$\left( \Delta \phi \right) _{brane}=\delta \left( \Delta \phi
\right) _{DM}=2\pi \delta v_{tg}^{2}$, where, for simplicity, we
have adopted again the isothermal sphere model for ''standard ''
dark matter. Moreover, we assume that the tangential velocity is
related to the baryonic mass by the (approximate) Keplerian value,
$v_{tg}^{2}\approx GM_{B}/r_{0}$. Therefore we obtain for the
tangential shear the following simple expression
\begin{equation}\label{pred}
\gamma _{\tan }\left( \theta \right) \approx 2\pi \delta
\frac{\left( \theta _{E}^{(GR)}\right) ^{2}}{\theta ^{2}}.
\end{equation}

This form of the tangential shear is similar to that one resulting
from a modification of the deflection law proposed by \cite{Ho02},
and which is given by $\gamma _{\tan }\left( \theta \right)
=\left( \theta _{out}/\theta _{0}\right)
\left( \theta _{E}^{2}/\theta ^{2}\right) $, where $\theta _{out}$ and $%
\theta _{0}$ are some ad hoc parameters. The allowable range for the values of $%
\theta _{0}$, obtained from observations, excludes small values.
The brane world model fixes the values of these parameters as
$\theta _{out}/\theta _{0}=2\pi \delta $. Since the
ratio $\theta _{out}/\theta _{0}$ is a measure of the ''mass discrepancy'', $%
\delta $ could also be interpreted as a measure of the extra mass
generated by the dark radiation. The best fit with the
observational data is obtained
for $\theta _{out}=s$, where $s$ is the truncation parameter. Therefore $%
\delta $ can also be expressed as $\delta =R_{U}/\theta _{0}$.

The mean shear signal can be measured in galaxy-galaxy lensing
observations. We compare the predictions of the brane world model,
given by Eq. (\ref{pred}) with the observational data obtained in
\citet{Ho04}. By fitting an isothermal sphere model to the
tangential shear at radii smaller than $2'$ (corresponding to a
maximum radius of $\approx 350h^{-1}$ kpc) the mean value for the
Einstein radius was found to be $\left<\theta
_E\right>=0.^{''}140\pm 0.^{''}009$. For this value of the
Einstein radius the variation of $\gamma _{\tan }$ calculated for
different values of the parameter $\delta $ is compared with the
observational data presented in \citet{Ho04}, in Fig. 3. The best
fit with the observational data corresponds to large values of
$\delta $, $\delta \approx 8-12$. Since the velocity dispersion
for the considered sample of galaxies is around $\sigma {\in
(128,150)}$ km/s, corresponding to $v_{tg}\approx 150-210$ km/s,
it follows that the ratio of the baryonic mass and the radius for
these galaxies should be of the order of $\chi \in (1.5\times
10^{-5},5\times 10^{-5})$. If the baryonic mass-radius ratio for
the galaxies is significantly smaller than this value, then the
predictions of this simple brane world model could not fit
satisfactorily the existing observational data. An estimate of the
galactic mass-radius ratio can be obtained by approximating the
tangential value by its Keplerian value, which gives $\chi \approx
8\sigma ^2$. For $\sigma =150$ km/s we obtain $\chi \approx
2\times 10^{-6}$. This simple estimate seems to indicate at first
sight that the present observational data do not favor the brane
world interpretation for dark matter. However, we have to mention
that the estimation of the ratio of the baryonic mass and the
radius from the tangential velocity could be affected by
significant errors. It would be more realistic to assume that
$\chi \approx 4k^2 v_{tg}^2=8k^2\sigma ^2$, where $k$ is a factor
describing the uncertainty in the determination of the ratio of
the "normal" mass and radius of a galaxy from the tangential
velocity alone. If this factor is of the order of $2$, $k\approx
2$, then the brane world model prediction could become consistent
with the observations. Hence to find a definite answer to the
possibility of the brane world models to correctly describe the
weak lensing data a much more careful analysis of data is needed.

Therefore the study of brane world model galaxy-galaxy lensing and
of the dark matter halos properties could provide strong
constraints on the brane world model and on related high energy
physics models.

The measurement of the azimuthally averaged tangential shear
around galaxies is robust against residual systematics, that is,
contributions from a constant or gradient residual shear cancel.
However, this is not the case for the quadrupole signal
\citep{Ho04}. If the lens galaxies are oriented randomly with
respect to the residual shear, then the average over many lenses
will cancel the contribution from systematics. But in real
observations the uncorrected shapes of the lenses are aligned with
the systematic signal. An imperfect correction can give rise to a
small quadrupole signal, which can mimic a cosmic shear signal.
Generally, the residual shear has an amplitude $\hat{\gamma}$ and
is located with respect to the major axis of the lens at an angle $\phi $. The tangential shear $%
\gamma _{\tan }^{obs}$ observed at a point $\left( r,\theta
\right) $ is the sum of the lensing signal $\gamma _{\tan
}^{lens}$ and the contribution from the systematics,
$\hat{\gamma}_{\tan }$, so that  $\gamma _{\tan }^{obs}=\gamma
_{\tan }^{lens}+\hat{\gamma}_{\tan }$. $\hat{\gamma}_{\tan }$ is
given by $\hat{\gamma}_{\tan }=-\hat{\gamma}\cos \left[ 2\left(
\theta -\phi \right) \right] $ \citep{Ho04}. One way to estimate
the flattening of the halo is
to measure the shears $\gamma _{\tan }^{(+)}$ at $\theta =0$ and $\pi $ and $%
\gamma _{\tan }^{(-)}$ at $\theta =\pi /2$ and $3\pi /2$,
respectively. The observed ratio is $f_{obs}=\gamma _{\tan
}^{(-)}/\gamma _{\tan }^{(+)}$ can be written as $f_{obs}=\left[
\gamma _{\tan }^{(-)}+\hat{\gamma}\cos \left( 2\phi \right)
\right] /\left[ \gamma _{\tan }^{(+)}-\hat{\gamma}\cos \left(
2\phi \right) \right] $ \citep{Ho04}. In the framework of the
simple brane world model we have considered all the shear
parameters (tangential as well as residual) are proportional to
the values corresponding to the standard dark matter case. On the
other hand the mean value of  $\cos \left( 2\phi \right) $ is
proportional to the measure $\alpha $ of the correlation between
the position angle of the lens and the direction of the systematic
shear of the background galaxies, which is a very small quantity.
The brane world model correction to $\alpha $ cannot lead to a
significant increase to this quantity, due to the large separation
distances between galaxies. Therefore we expect that $f_{obs}$ is
an invariant and brane world effects do not affect the robustness
of the measurement of the average halo shape.

With the use of the approximate relation $
 \left(1+x/n\right) ^{n}\approx \exp \left( x\right)$,
which is valid for large $n$, we can write Eq. (\ref{condr}) in the form
\begin{equation}
R_{b}\approx r_{0}\exp \left( \chi /4v_{tg}^{2}\right).
\end{equation}

In the case of a very massive object with $\chi \approx 1$ and for
$v_{tg}\in \left( 10^{-4},10^{-3}\right) $ it follows that
$R_{b}>>r_{0}$. But for very small values of $\chi $, of the same
order as $v_{tg}^{2}$, $R_{b}\approx r_{0}$. Therefore flat
rotation curves are specific for particles moving in stable
circular orbits around galaxies (having $\chi <<1$), while the
same phenomenon cannot be detected for very compact (stellar or
black hole type) objects.

The measurement of the anisotropy in the lensing signal around
galaxies and the detection of the flattening of the dark matter
halos could pose some serious challenges to the alternative
theories of gravitation  \citep{Ho02,Ho04}. In many theories
proposing modifications of the dynamic law for gravitation the
lensing signal caused by the intrinsic shapes of the galaxies
decreases as $\propto r^{-2}$ and hence it is negligible at large
distances. Therefore dark halos cannot be modelled as spherical
systems. This important observational result can clearly
discriminate between the gravitational explanations proposed as
substitutes for dark matter. However, in the framework of the
model discussed in the present paper, the flattening of the dark
matter halos and the anisotropic signal can be easily explained,
at least at a simple qualitative level. We have obtained all our
results under the (unrealistic) assumption of the spherical
symmetry, with the galaxy (consisting of baryonic matter only)
modelled as a self-gravitating relativistic sphere. A much more
realistic model for a galaxy is its representation as a
self-gravitating rotating stationary axisymmetric disc, in which
generally the metric coefficients depend on both polar coordinates $r$ and $%
\theta $. The brane world metric outside this baryonic matter
distribution is also anisotropic, and we would naturally expect a
flattening of the region in which the bulk effects are important,
associated with an anisotropic (angular dependent) weak lensing
signal. The ''dark matter'' is, in this model, a projection of the
geometry of the higher dimensional space time far away from the
baryonic matter distribution, which has to match the geometry of
the galaxy. Therefore disc type or spheroidal distributions will
automatically create flattened and anisotropic (non-spherical)
geometrical effects. For example, already in standard general
relativity in the weak field limit a cold disc has the associated
gravitational potential $-\nu
=\Phi =-v_{tg}^{2}\ln \left[ r\left( 1+\left| \cos \theta \right| \right) /D%
\right] $, where $D$ is a fiducial length scale. The bending of
light is anisotropic (polar angle dependent), as is the focussing
effect of the disc \citep{Ca02}. Therefore, more realistic
geometric distributions of the baryonic matter inside the galaxy
and of the exterior space-time geometry of the brane can easily
explain the flattening effect and the anisotropic lensing observed
for dark matter halos.

One of the main advantages of the brane world models, as compared
to the other alternative explanations of the galactic rotation
curves and of the dark matter, is that it can give a systematic
and coherent description of the Universe from galactic to
cosmological scales. On the other hand, in the present model all
the relevant physical quantities, including the dark energy and
the dark pressure, which describe the non-local effects due to the
gravitational field of the bulk, are expressed in terms of
observable parameters (the baryonic mass, the radius of the galaxy
and the observed flat rotational velocity curves). Therefore this
opens the possibility of testing the brane world models by using
astronomical and astrophysical observations at the galactic scale.
In this paper we have provided some basic theoretical tools
necessary for the in depth comparison of the predictions of the
brane world model and of the observational results.

\section*{Acknowledgements}

We are grateful to the anonymous referee whose comments and
suggestions have significantly improved the quality of the paper.
This work is supported by a RGC grant of the government of the
Hong Kong SAR.

\clearpage

\begin{figure}
\plotone{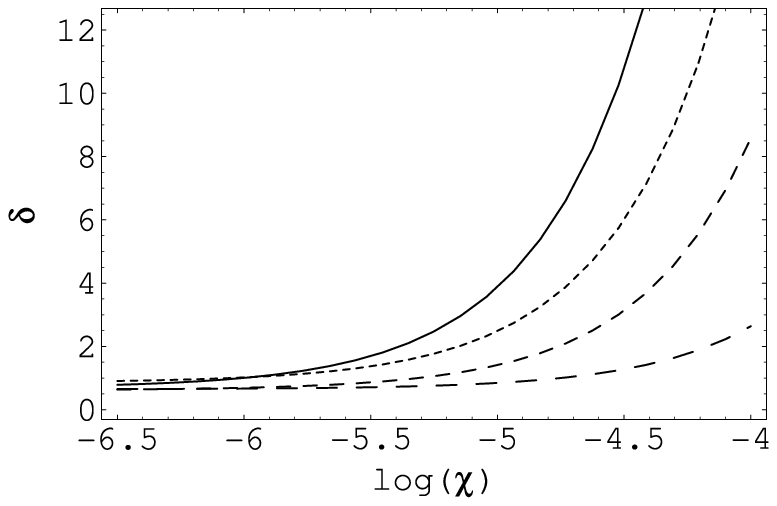} \caption{The ratio $\delta =\left(\Delta \phi
\right)_{brane}/2\pi v_{tg}^2$ of the light deflection angle in
the galactic metric on the brane and the deflection angle in the
semi-realistic isothermal galactic dark matter model as a function
of the parameter $\chi =4GM_B/r_0$ for different values of the
tangential velocity: $v_{tg}=150$ km/s (solid curve), $v_{tg}=210$
km/s (dotted curve), $v_{tg}=300$ km/s (short dashed curve) and
$v_{tg}=450$ km/s (long dashed curve). For the impact parameter
$r_0$ and the cosmological constant $\Lambda $ we have adopted the
values $r_0=30$ kpc and $\Lambda =3\times 10^{-56}$ cm$^{-2}$,
respectively. \label{FIG1}}
\end{figure}

\begin{figure}
\plotone{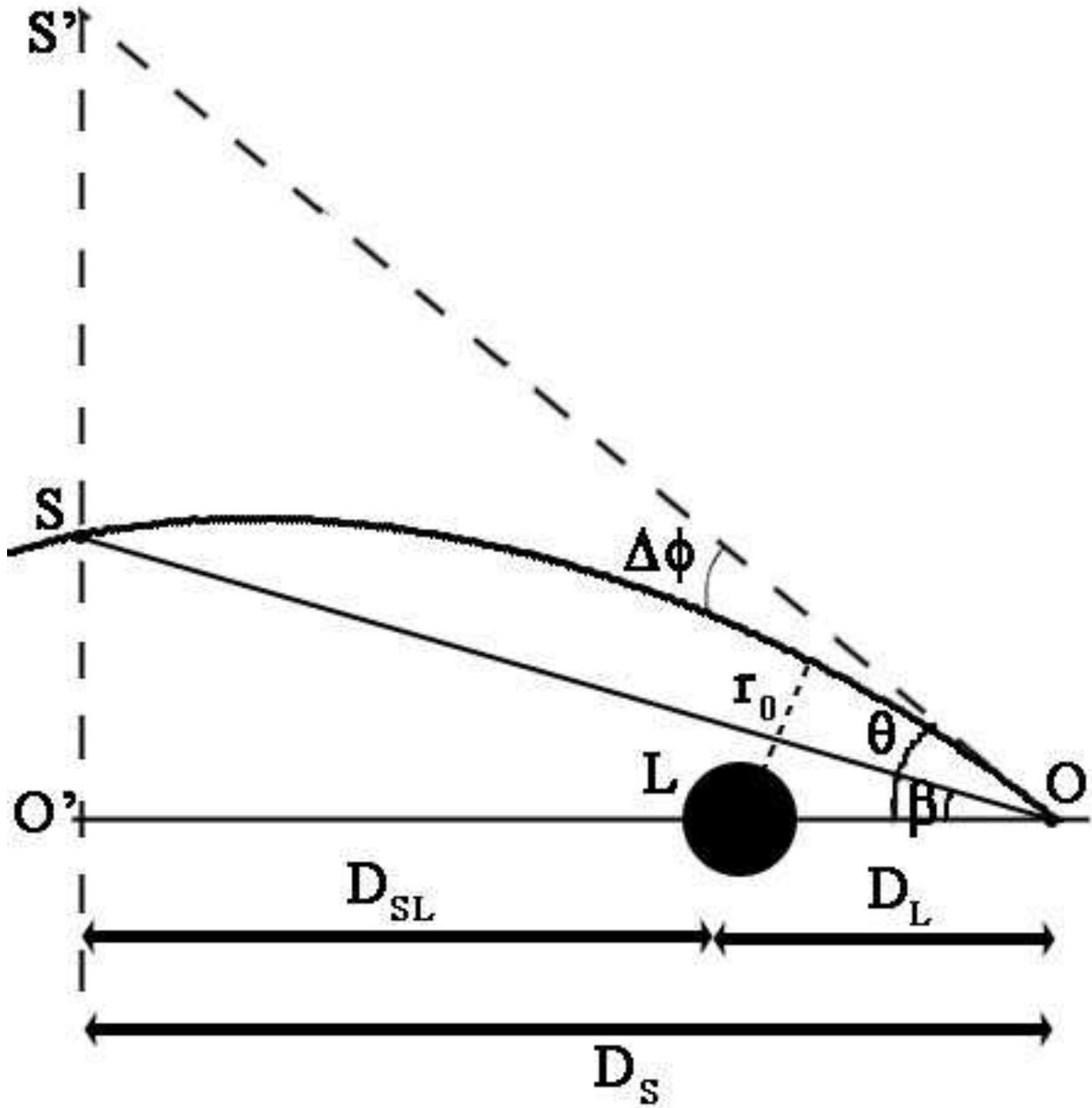} \caption{The lensing geometry, showing the
location of the observer $O$, the lensing galaxy $L$ and the
source $S$. The deflection angle is $\Delta \phi$. The angular
diameter distances $D_L$, $D_{LS}$ and $D_S$ are also indicated.
\label{FIG2}}
\end{figure}

\begin{figure}
\plotone{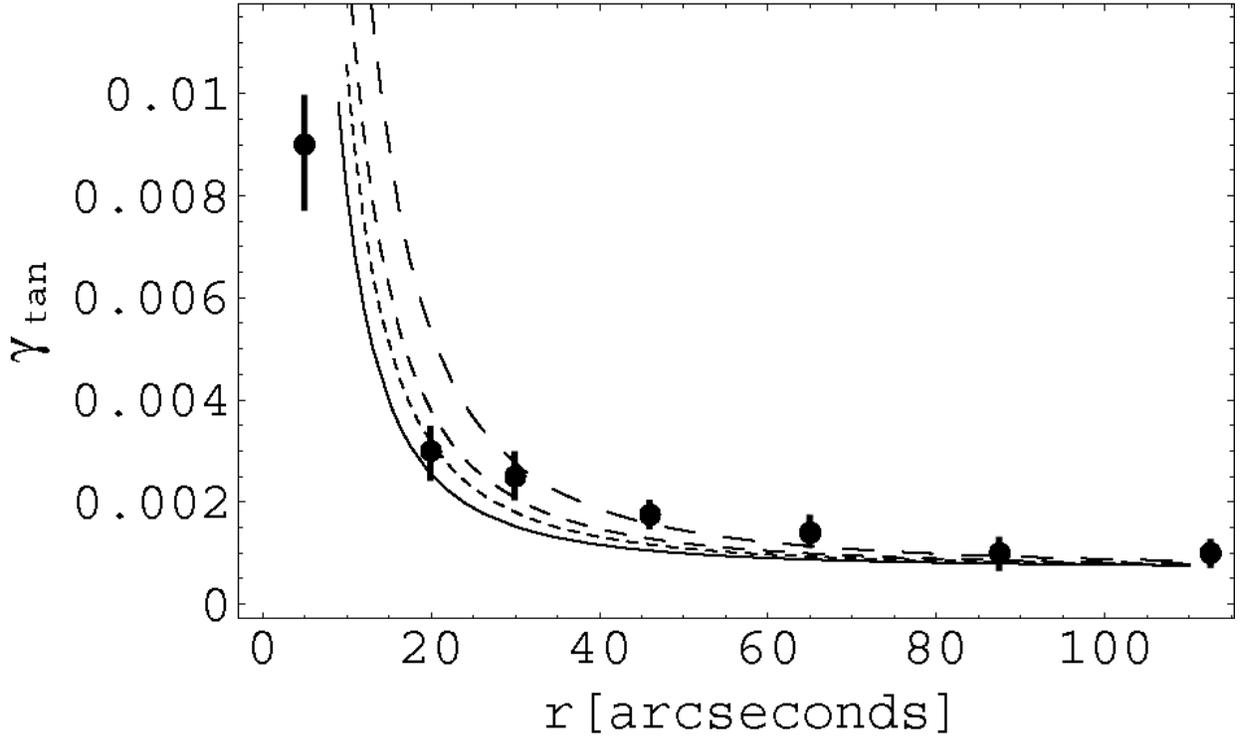} \caption{Comparison of the variation of the
tangential shear as a function of the distance in the brane world
model and the observational data of \citet{Ho04} (represented by
points) for different values of the parameter $\delta $: $\delta
=6$ (solid curve), $\delta =8$ (dotted curve), $\delta =10$
(dashed curve), $\delta =12$ (long dashed curve). The Einstein
radius is $\theta _E=0.^{''}14$. \label{FIG3}}
\end{figure}

\end{document}